\documentclass[a4paper, 11pt]{article}  
\usepackage{geometry, graphicx, blindtext, authblk}     
\usepackage{amssymb, amsmath}

\usepackage[table]{xcolor}
\usepackage[colorlinks=true, linkcolor=blue, citecolor = blue]{hyperref}
\usepackage{cleveref} 
\usepackage{booktabs, color}
\usepackage{tikz}
\usetikzlibrary{arrows,shapes,trees} 
\usetikzlibrary{arrows.meta}
\newcommand{\bx}{\mathbf{x}}
\newcommand{\by}{\mathbf{y}}
\newcommand{\bh}{\mathbf{h}}
\newcommand{\bC}{\mathbf{C}}
\newcommand{\bc}{\mathbf{c}}

\title{Kriging and Gaussian processes: interconnected frameworks for spatial modelling and prediction}
\author{Hossein Mohammadi\thanks{Corresponding Author: h.mohammadi@exeter.ac.uk}}
\affil{Faculty of Environment, Science and Economy, University of Exeter, UK}
\date{}							
\begin{document}
\maketitle
\begin{abstract}
Spatial modelling and inference arise across many areas of science and engineering, with Kriging and Gaussian process (GP) regression serving as two widely used predictive models. Although grounded in the same random field theory, these approaches are often treated in isolation. As a result, their connection has been obscured by differences in terminology, notation, and practical implementation. This separation largely reflects their independent development within geostatistics and machine learning, respectively. To bridge this methodological divide, this work establishes a unified mathematical framework that connects the two paradigms. The underlying assumptions and formulations of each approach are detailed and compared within this framework. This comparison highlights their differences while clarifying the connections between them. The resulting perspective provides practitioners with a common foundation for translating concepts and methods between geostatistics and machine learning. In particular, it makes methodological developments in either field more accessible to the other, including advances in uncertainty quantification from the GP literature.
\end{abstract}
\section{Introduction}
\label{sec:introduction}
%===================================================================================================
Many disciplines in science and engineering involve the study of complex physical systems that exhibit spatial variation. Examples include reservoir modelling in geostatistics \cite{pyrcz2014}, pollutant tracking in atmospheric science \cite{kim2025}, remote sensing analysis of satellite images \cite{roy2017}, and gene expression profiling in spatial transcriptomics \cite{sun2020}. Since measuring such systems continuously across space is impossible due to physical, financial, and logistical constraints, the primary objective is to predict values at unobserved locations using data collected from a finite set of sampled sites. 

Accurate spatial prediction models are therefore essential for these applications. The choice of predictive framework depends on the type of spatial data and its underlying structure. Geostatistical data, lattice (or areal) data, and spatial point patterns are the three main categories of spatial data identified by \cite{cressie1993}. A detailed discussion of the distinctions between these categories can be found in \cite{moraga2023}. 

This work focuses exclusively on geostatistical data, where a quantity of interest is measured at a finite set of locations within a continuous domain. In this setting, Kriging \cite{cressie1993, stein1999} has long been established as a standard method for optimal interpolation, originating from the empirical work of the South African mining engineer Danie G. Krige \cite{krige1953}. This empirical approach was later formalised and developed in the 1960s by the French mathematician Georges Matheron, who established the theoretical basis of modern geostatistics \cite{matheron1963}. His formulation relies on random field theory \cite{stein1999, hristopulos2020}, detailed in the next section.

The underlying principles of Kriging are deeply interconnected with GP regression \cite{GPML}, a non-parametric approach that places a probability distribution over functions. However, they emerged from different scientific disciplines, namely, geostatistics and machine learning, respectively. For example, Kriging is derived under the Best Linear Unbiased Prediction (BLUP) criterion, whereas GPs adopt a probabilistic paradigm that provides a full predictive distribution. The two methodologies also differ in how spatial dependence is represented. Kriging uses the semivariogram to quantify dissimilarity between observations, while GP regression uses the covariance function to quantify similarity.

Furthermore, Kriging has traditionally been applied to low-dimensional problems in two or three dimensions, where the inputs represent geographic locations. GP regression, however, extends beyond conventional spatial settings and can accommodate input spaces of higher dimension \cite{GPML}. Today, GPs are frequently used as surrogate models for computationally expensive black-box functions \cite{gramacy2020}. Applications include the design and analysis of computer experiments \cite{santner2018}, Bayesian optimisation \cite{shahriari2016}, robotic motion planning and control \cite{cosier2024}, and the numerical solution of partial differential equations \cite{chen2021}.

This historical divide often obscures the theoretical connections between the two paradigms through differences in terminology, notation, and practical implementation. Moreover, since the literature largely treats these approaches in isolation, methodological advances in one community rarely inform the other. The present paper addresses this divide via a unified mathematical framework that connects Kriging and GPs. To do so, we introduce and analyse the foundational assumptions and formulations of both approaches and clarify where they differ. This unified perspective is particularly valuable for practitioners seeking to transfer software tools, techniques, and computational methods between the two domains, including techniques for uncertainty quantification using GPs. Several studies have investigated the connection between Kriging and GP regression (see, e.g., \cite{suryasentana2023,marinescu2024}). Building on this literature, the present paper provides a more comprehensive treatment of the two approaches, with an emphasis on their mathematical foundations. 

The remainder of this paper is structured as follows. \Cref{sec:RFs,sec:RF_properties} present the theoretical foundations by introducing random fields and detailing their statistical properties. \Cref{sec:Kriging,sec:GP_regression} focus on spatial prediction with Kriging and GP regression, respectively. \Cref{sec:variogram_covariance_fun} provides a deeper analysis of covariance and variogram functions, which leads directly into the treatment of noisy observations and the nugget effect in \Cref{sec:noisy_RF}. \Cref{sec:param_estim} addresses the computational methods for estimating unknown model parameters, and \Cref{sec:conclusion} concludes the report with a summary of key insights.
%===================================================================================================
\section{Random fields}
\label{sec:RFs}
%===================================================================================================
Classical statistical methods, such as linear regression, generally assume that residuals are independent and identically distributed. Spatial data, however, often violate the independence assumption, as observations collected at nearby locations tend to be more similar than those measured further apart \cite{tobler1970}. This is due to the fact that observations are often influenced by common underlying physical processes. Random fields (RFs) provide a principled way to account for this spatial dependence.

Consider a probability space $(\Omega,\mathcal{F},\mathbb{P})$, where $\Omega$ denotes the sample space, $\mathcal{F}$ a $\sigma$-algebra, and $\mathbb{P}$ a probability measure. Suppose that the underlying spatial process is defined over a continuous domain $\mathcal{X} \subset \mathbb{R}^d$. In geographical applications, $\mathcal{X}$ typically represents spatial coordinates (such as longitude and latitude), so $d = 2$. An RF, denoted by $\{Y(\bx, \omega): \bx \in \mathcal{X}\}$, is a mapping $Y: \mathcal{X} \times \Omega \to \mathbb{R}$ such that:
\begin{itemize}
	\item for each $\bx \in \mathcal{X}$, the function $Y(\bx,\cdot)$ is a random variable on
	$(\Omega,\mathcal{F},\mathbb{P})$;
	\item for each $\omega \in \Omega$, the function $Y(\cdot,\omega)$ is a realisation of the field, called a \emph{sample path}.
\end{itemize}
Equivalently, an RF can be viewed as a collection of random variables indexed by the set $\mathcal{X}$. From this perspective, an RF is a stochastic process in which the usual one-dimensional index parameter 
(such as ``time'') is replaced by a higher-dimensional spatial domain or, more generally, a manifold \cite{adler2007}.

As with any stochastic process, an RF $Y(\bx, \omega)$ can be decomposed into a deterministic trend and a zero-mean residual RF:
\begin{equation}
	Y(\bx, \omega) = \mu(\bx) + Z(\bx, \omega). 
\end{equation}
Here, $\mu(\bx) = \mathbb{E}[Y(\bx, \omega)]$ is the expectation of the field, referred to as the \emph{trend function}, and $Z(\bx, \omega)$ is a zero-mean RF capturing local fluctuations about the trend. For brevity, we suppress the realisation argument $\omega$ hereafter and write $Y(\bx)$.

 Let $Y(\bx)$ be a second-order RF, so that $\mathbb{E}[Y(\bx)^2] < \infty$ for all $\bx\in \mathcal{X}$. The spatial dependence of $Y(\bx)$ is typically quantified via covariance or variogram functions. The covariance function (also known as the kernel), denoted by $C$, characterises the spatial dependence between values at any two locations $\bx, \bx'\in \mathcal{X}$, and is defined as:
\begin{equation}
	C(\bx, \bx') = \mathbb{E}\bigl[\bigl(Y(\bx)-\mu(\bx)\bigr)\bigl(Y(\bx')-\mu(\bx')\bigr)\bigr].
\end{equation}
In contrast, the variogram quantifies spatial dissimilarity through the variance of the difference between field values at two locations:
\begin{equation}
	2\gamma(\bx, \bx')
	= \operatorname{Var}\bigl[Y(\bx) - Y(\bx')\bigr].
	\label{eq:variogram}
\end{equation}
Half of this quantity, $\gamma(\bx, \bx')$, is defined as the \emph{semivariogram}. These functions are discussed further in \Cref{sec:variogram_covariance_fun}.
%===================================================================================================
\section{Properties of random fields}
\label{sec:RF_properties}
%===================================================================================================
An RF is characterised by its finite-dimensional distributions. Specifically, for any $n\in\mathbb{N}$ and any set of locations $\bx_1,\ldots,\bx_n\in \mathcal{X}$, consider the random vector \begin{equation} 
	\by = \bigl(Y(\bx_1),\ldots,Y(\bx_n)\bigr)^\top. 
	\label{joint_vector} 
\end{equation} 
The complete family of these distributions uniquely determines the probability law of the process.

\begin{figure}[htbp] 
	\centering
	\includegraphics[width=0.49\textwidth]{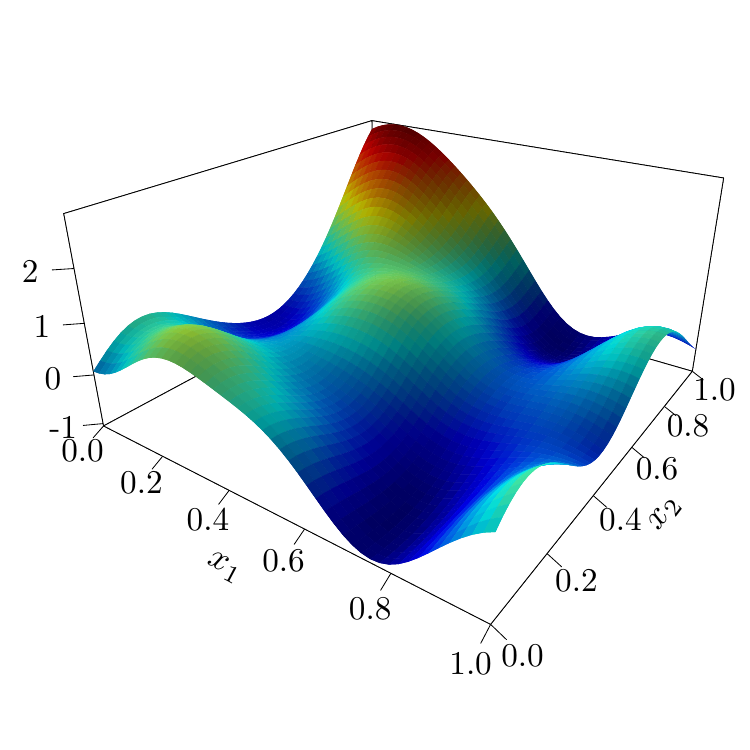}
	\includegraphics[width=0.49\textwidth]{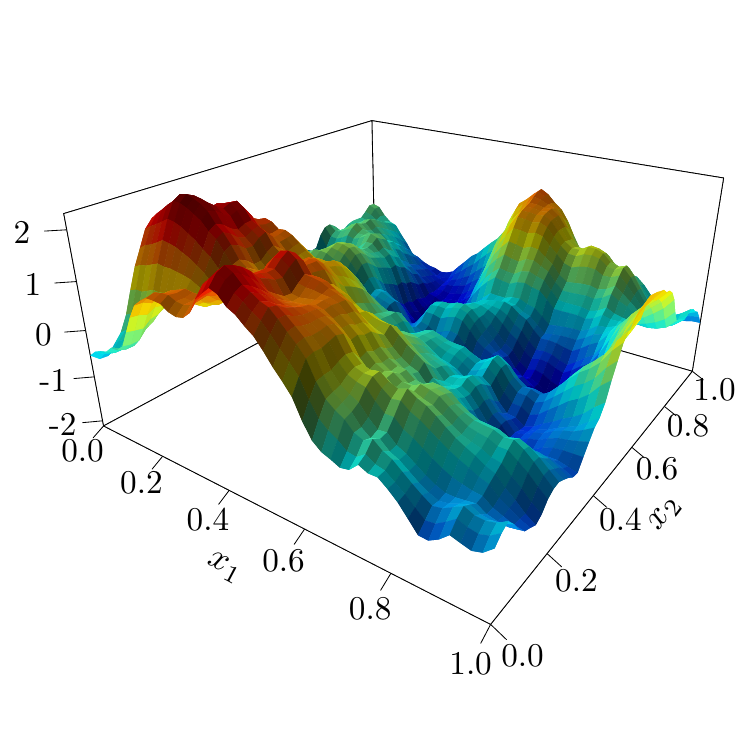}  
	\caption{Two realisations from zero-mean GRFs over the domain $\mathcal{X}=[0,1]^2$. The fields exhibit different patterns of spatial dependence due to their distinct covariance structures.}
	\label{fig:gp_realisation}
\end{figure}

A particularly important class of RFs is the \emph{Gaussian random field} (GRF) \cite{haran2011}, for which $\by$ in \eqref{joint_vector} follows a multivariate Gaussian distribution. Since Gaussian distributions are uniquely determined by their first two moments, a GRF is fully specified by its mean and covariance functions. \Cref{fig:gp_realisation} illustrates two realisations from GRFs with different covariance structures, both with zero mean ($\mu(\bx)=0$).

We now introduce several properties of RFs that are fundamental to spatial modelling and inference. Throughout this paper, $\bh\in\mathbb{R}^d$ denotes the spatial lag (or separation) vector.

\paragraph{Strict stationarity}
An RF is strictly (or strongly) stationary if its finite-dimensional distributions are invariant to spatial translation. More precisely, 
for any $n\in\mathbb{N}$, any locations $\bx_1,\ldots,\bx_n\in \mathcal{X}$, and any spatial 
lag vector $\bh$ such that $\bx_1+\bh,\ldots,\bx_n+\bh\in \mathcal{X}$, the 
corresponding random vectors are equal in distribution:
\[
\bigl(Y(\bx_1),\ldots,Y(\bx_n)\bigr)^\top
\stackrel{\mathrm{dist}}{=}
\bigl(Y(\bx_1+\bh),\ldots,Y(\bx_n+\bh)\bigr)^\top.
\]
Under strict stationarity, the probability law of the process is invariant to translation: shifting any set of coordinates by a vector $\bh$ leaves its joint distribution unchanged. 

\paragraph{Second-order stationarity}
The most common framework in spatial modelling is \emph{second-order} (or \emph{weak}) stationarity. Unlike the strict condition, it is defined entirely in terms of the first two moments. An RF $Y(\bx)$ satisfies this condition if:
\begin{enumerate}
	\item The mean function is constant across the domain, i.e.,
	\[
	\mu(\bx) = \mu_0, \qquad \forall\,\bx\in \mathcal{X};
	\]
	\item The covariance between any two values is determined only by their spatial lag vector $\bh$, such that
	\[
	\operatorname{Cov}\bigl[Y(\bx), Y(\bx + \bh)\bigr] = C(\bh), \qquad \forall\,\bx, \bx+\bh\in \mathcal{X}.
	\]
\end{enumerate}
In other words, the spatial covariance depends exclusively on relative separation rather than absolute position. An immediate consequence of this property is that the variance across the field is constant and equal to the covariance at zero lag:
\[
\sigma_0^2 = \operatorname{Var}\bigl[Y(\bx)\bigr] = C(\mathbf{0}), \qquad \forall\,\bx\in \mathcal{X}.
\]

Under second-order stationarity, expanding the variogram defined in \eqref{eq:variogram} yields
\begin{align*}
	2\gamma(\bh)
	&= \operatorname{Var}\bigl[Y(\bx+\bh) - Y(\bx)\bigr] \\
	&= \operatorname{Var}\bigl[Y(\bx+\bh)\bigr] + \operatorname{Var}\bigl[Y(\bx)\bigr] - 2\operatorname{Cov}\bigl[Y(\bx+\bh), Y(\bx)\bigr] \\
	&= \sigma_0^2 + \sigma_0^2 - 2C(\bh) \\
	&= 2\bigl(C(\mathbf{0}) - C(\bh)\bigr).
\end{align*}
Dividing by two establishes the fundamental relationship between the semivariogram and covariance function:
\begin{equation}
	\gamma(\bh) = C(\mathbf{0}) - C(\bh).
	\label{variog_cov_relation}
\end{equation}

It is important to note that a strictly stationary field is not necessarily second-order stationary. Strict stationarity requires translation invariance of the finite-dimensional distributions but does not require the existence of statistical moments. Thus, an RF can satisfy distributional invariance while its mean or variance remains undefined. For example, a \emph{Cauchy RF} may have translation-invariant multivariate Cauchy finite-dimensional distributions and hence be strictly stationary. However, since its first and second moments do not exist, it fails to satisfy the second-order criterion. Conversely, a strictly stationary process with finite second moments is necessarily second-order stationary.

\paragraph{Isotropy}
A common additional assumption under second-order stationarity is \emph{isotropy}, whereby the covariance function depends solely on the Euclidean distance $\|\bh\|$ rather than the direction of the vector $\bh$:
\[
\operatorname{Cov}\bigl[Y(\bx), Y(\bx + \bh)\bigr] = C(\|\bh\|), \qquad \forall\,\bx, \bx+\bh\in \mathcal{X}.
\]
This property expresses the rotational invariance of the spatial structure: pairs of locations separated by the same distance share the same covariance, regardless of the direction of separation. The fields displayed in \Cref{fig:gp_realisation} are realisations of isotropic GRFs.

\begin{figure}[htbp] 
	\centering
	\includegraphics[width=0.49\textwidth]{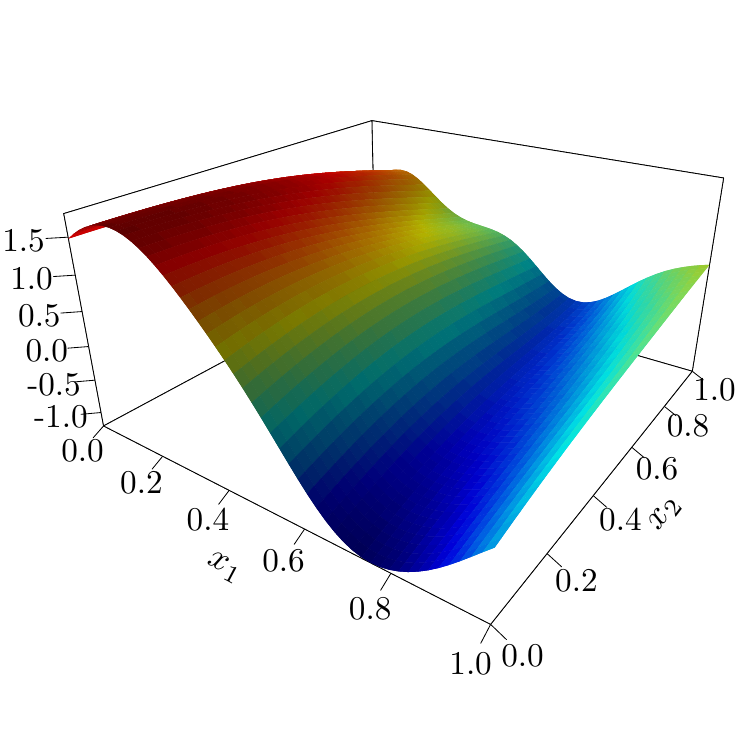} 
	\includegraphics[width=0.49\textwidth]{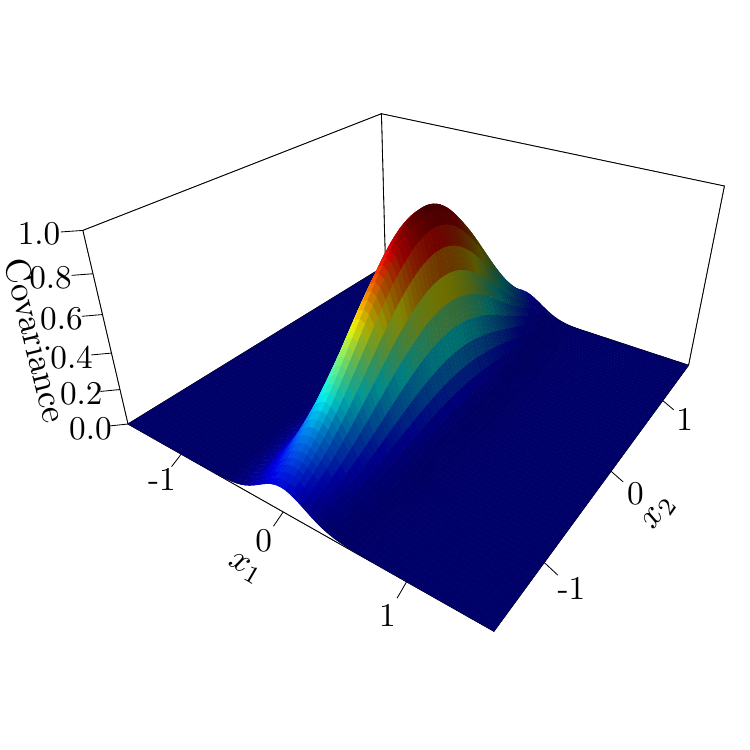} 
	\caption{Realisation of an anisotropic GRF (left) and its corresponding covariance function (right). The covariance decays more rapidly along $x_1$ than along $x_2$, resulting in more rapid spatial variation along $x_1$.}
\label{fig:aniso_gp}
\end{figure}

By contrast, a field is said to be \emph{anisotropic} if its covariance cannot be expressed as a function of $\|\bh\|$ alone. In other words, the covariance structure of an anisotropic field varies with direction as well as distance. \Cref{fig:aniso_gp} illustrates a realisation of an anisotropic GRF alongside its corresponding covariance function. Here, the covariance decays more rapidly along $x_1$ than along $x_2$; consequently, the field varies more rapidly along the first coordinate axis than along the second.

\paragraph{Intrinsic stationarity} 
Introduced by Matheron \cite{matheron1973}, intrinsic stationarity focuses on the behaviour of the field's \emph{increments}, $Y(\bx+\bh) - Y(\bx)$. Unlike second-order stationarity, it does not require the variance of the field itself to exist, making it a weaker and more general notion. Formally, an RF is intrinsically stationary if the following conditions hold for all $\bx, \bx+\bh \in \mathcal{X}$:
\begin{enumerate}
	\item The expected value of the increment is zero, i.e.,
	\[
	\mathbb{E}\bigl[Y(\bx+\bh) - Y(\bx)\bigr] = 0\,;
	\]
	\item The variance of the increment exists and depends solely on the lag vector $\bh$, such that
	\[
	\operatorname{Var}\bigl[Y(\bx+\bh) - Y(\bx)\bigr] = 2\gamma(\bh)\,,
	\]
	where $\gamma(\bh)$ denotes the semivariogram function.
\end{enumerate}
In other words, intrinsic stationarity requires translation invariance of the first two moments of the increments rather than those of the field itself.

A second-order stationary field is always intrinsically stationary, as its constant mean ensures that the expected value of any increment is zero. The second condition also follows directly from \eqref{variog_cov_relation}, since the variance of the increments depends only on the lag vector $\bh$. However, for an intrinsically stationary RF, the variance of the field itself need not exist, meaning that it may fail to satisfy second-order stationarity.

\paragraph{Separability} 
An RF is said to have a \emph{separable} structure if its spatial dependence can be written as the product of one-dimensional covariance functions, one for each coordinate. Formally, a covariance function $C$ is separable if
\begin{equation}
	C(\bx, \bx') = \prod_{i=1}^{d} C_i(x_i, x'_i),
\end{equation}
where $C_i$ denotes the univariate covariance function associated with the $i$-th coordinate, and $x_i$ and $x'_i$ are the $i$-th components of $\bx$ and $\bx'$, respectively.

Separable structures are computationally attractive because they allow the covariance matrix to be expressed as a Kronecker product of smaller matrices. This can considerably reduce both storage and computational costs; see \Cref{sec:param_estim_GPs} for further details. However, separability may limit modelling flexibility, as the dependence along one dimension cannot vary with another. For example, a separable space–time model cannot capture spatial dependence that changes over time.

\paragraph{Continuity} 
The continuity of an RF can be studied in various ways; see \cite{hristopulos2020} for a comprehensive review. Here, we focus on continuity in the mean-square (MS) sense. An RF $Y(\bx)$ is said to be MS continuous at a location $\bx_0 \in \mathcal{X}$ if 
\begin{equation} 
	\lim_{\bx \to \bx_0} \mathbb{E}\Bigl[\bigl(Y(\bx)-Y(\bx_0)\bigr)^2\Bigr] = 0. 
	\label{MS_continuity} 
\end{equation} 
If this condition holds for every $\bx_0 \in \mathcal{X}$, the field is MS continuous on its domain $\mathcal{X}$. 

Under second-order stationarity, the mean is constant and the covariance function depends only on the separation vector $\bh$. The expectation in \eqref{MS_continuity} can be written as
\[ 
\mathbb{E}\Bigl[\bigl(Y(\bx)-Y(\bx_0)\bigr)^2\Bigr] = 2\sigma_0^2 - 2C(\bx-\bx_0). 
\]
As $\bh=\bx-\bx_0 \to \boldsymbol{0}$, this quantity converges to zero if and only if 
\[ 
\lim_{\bh\to\boldsymbol{0}} C(\bh) = C(\boldsymbol{0}) = \sigma_0^2. 
\]
Thus, a second-order stationary RF is MS continuous if and only if its covariance function is continuous at the origin. An important consequence of second-order stationarity is that MS continuity is a global property: if the field is MS continuous at one location, it is MS continuous throughout its domain.

\paragraph{Differentiability}
The smoothness of an RF can be further characterised in terms of MS differentiability. An RF $Y(\bx)$ is said to be MS differentiable at $\bx_0$ if, for each coordinate $i = 1, \dots, d$, there exists a random variable $\frac{\partial Y}{\partial x_i}(\bx_0)$ such that 
\begin{equation} 
	\lim_{\Delta x_i \to 0} \mathbb{E}\Biggl[ \biggl( \frac{Y(\bx_0 + \Delta x_i \mathbf{e}_i) - Y(\bx_0)} {\Delta x_i} - \frac{\partial Y}{\partial x_i}(\bx_0) \biggr)^2 \Biggr] = 0,
	\label{MS_differentiability} 
\end{equation} 
where $\mathbf{e}_i$ denotes the $i$-th standard basis vector. The field is said to be MS differentiable over its domain $\mathcal{X}$ if \eqref{MS_differentiability} holds for every $\bx_0 \in \mathcal{X}$ and every coordinate $i$. 

Under second-order stationarity, this condition simplifies further: the partial derivative $\frac{\partial Y}{\partial x_i}(\bx_0)$ exists in the MS sense if and only if the second-order partial derivative $\frac{\partial^2 C(\bh)}{\partial h_i^2}$ exists and is finite at the origin, $\bh = \boldsymbol{0}$. More generally, $Y(\bx)$ is $r$-times partially MS differentiable along the $i$-th coordinate axis if and only if $C(\bh)$ is $2r$-times partially differentiable with respect to $h_i$ at $\bh = \boldsymbol{0}$ \cite{stein1999}. MS differentiability is therefore global under second-order stationarity: with respect to each coordinate direction, if it holds at one location, it holds throughout the domain \cite{stein1999}.
%===================================================================================================
\section{Spatial interpolation with Kriging}
\label{sec:Kriging}
%===================================================================================================
This section introduces spatial interpolation using Kriging from a geostatistical perspective. The mean of the underlying RF plays a critical role in the formulation of Kriging. Depending on how the mean function $\mu(\bx)$ is specified, three classical variants of Kriging can be distinguished:
\begin{itemize}
	\item \textbf{Simple Kriging (SK):} $\mu(\bx) = \mu_0$ is known and constant across the domain and must be specified a priori.
	
	\item \textbf{Ordinary Kriging (OK):} $\mu(\bx)= \mu_0$ is constant but unknown. Its explicit estimation is not required to derive the OK equation, as explained below.
	
	\item \textbf{Universal Kriging (UK):} $\mu(\bx)$ varies spatially and is typically modelled as a linear combination of known basis functions. UK is beyond the scope of this study, which focuses exclusively on SK and OK.
\end{itemize}

Now suppose that the RF $Y(\bx)$ is sampled at $n$ locations $\bx_1,\ldots,\bx_n \in \mathcal{X}$. We wish to predict the value of $Y(\bx)$ at an unobserved target location $\bx_0$ using OK and SK, with the corresponding predictors denoted by $\hat{Y}_{\mathrm{ok}}(\bx_0)$ and $\hat{Y}_{\mathrm{sk}}(\bx_0)$, respectively. These predictors are derived under the \emph{Best Linear Unbiased Predictor} (BLUP) criterion.

The linearity requirement of BLUP implies that the predictors are constructed as linear combinations of the RF values. Accordingly, $\hat{Y}_{\mathrm{ok}}(\bx_0)$ and $\hat{Y}_{\mathrm{sk}}(\bx_0)$ are expressed as
\begin{align}
	\label{ok_predictor}
	\hat{Y}_{\mathrm{ok}}(\bx_0) &= \boldsymbol{\lambda}_{\mathrm{ok}}^{\top}\by,\\
	\hat{Y}_{\mathrm{sk}}(\bx_0) &= \mu_0 + \boldsymbol{\lambda}_{\mathrm{sk}}^{\top}(\by - \mu_0 \mathbf{1}),
	\label{SK_predictor}
\end{align}
where $\boldsymbol{\lambda}_{\mathrm{ok}}= [\lambda_{\mathrm{ok}, 1}, \ldots, \lambda_{\mathrm{ok}, n}]^{\top}$ and $\boldsymbol{\lambda}_{\mathrm{sk}} = [\lambda_{\mathrm{sk}, 1}, \ldots, \lambda_{\mathrm{sk}, n}]^{\top}$ are the weight vectors to be determined, and $\mathbf{1}$ denotes an $n \times 1$ vector of ones. The main difference between these two formulations is that $\hat{Y}_{\mathrm{sk}}(\bx_0)$ explicitly incorporates the known constant mean $\mu_0$. In practice, $\by$ is replaced by the vector of observed values 
\begin{equation}
	\by_{\mathrm{obs}} = [y(\bx_1), \ldots, y(\bx_n)]^{\top} .
\end{equation}

The unbiasedness property requires that the expected prediction error be zero or, equivalently, that the expected value of the predictor equals that of the underlying RF. For a linear candidate predictor $\hat{Y}(\bx_0)$, this criterion is expressed as
\[
\mathbb{E}\bigl[\hat{Y}(\bx_0) - Y(\bx_0)\bigr] = 0 \iff \mathbb{E}\bigl[\hat{Y}(\bx_0)\bigr] = \mathbb{E}\bigl[Y(\bx_0)\bigr]\,.
\]
For OK, evaluating the expectation of the prediction error yields
\begin{align}
	\mathbb{E}\bigl[\hat{Y}_{\mathrm{ok}}(\bx_0) - Y(\bx_0)\bigr]
	&= \mathbb{E}\biggl[\boldsymbol{\lambda}_{\mathrm{ok}}^{\top}\by - Y(\bx_0)\biggr] \nonumber \\
	&= \mu_0 \boldsymbol{\lambda}_{\mathrm{ok}}^{\top} \mathbf{1} - \mu_0 \nonumber \\
	&= \mu_0\biggl(\boldsymbol{\lambda}_{\mathrm{ok}}^{\top} \mathbf{1} - 1\biggr).
\end{align}
For the prediction error to have zero expectation for any value of the unknown mean $\mu_0$, the weights must satisfy
\begin{equation}
	\boldsymbol{\lambda}_{\mathrm{ok}}^{\top} \mathbf{1} = \sum_{i=1}^{n}\lambda_{\mathrm{ok}, i} = 1.
\end{equation}
This requirement defines the OK unbiasedness constraint, which holds independently of the value of $\mu_0$.  Hence, $\mu_0$ does not need to be estimated explicitly. By contrast, the SK predictor inherently satisfies the unbiasedness condition for any choice of weights, since
 \begin{equation}
 	 \mathbb{E}[\hat{Y}_{\mathrm{sk}}(\bx_0)] = \mu_0 + \boldsymbol{\lambda}_{\mathrm{sk}}^{\top}\bigl(\mu_0 \mathbf{1} - \mu_0 \mathbf{1}\bigr) = \mu_0 = \mathbb{E}[Y(\bx_0)]. 
\end{equation}
Consequently, the SK weights are unconstrained.

Finally, the optimal (or ``best'') property in BLUP signifies that among all linear and unbiased predictors, the Kriging predictor minimises the prediction error variance. Since the predictor is unbiased, this variance equals the mean squared prediction error (MSPE):
\begin{equation}
	\mathrm{MSPE}(\bx_0) = \mathbb{E}\Bigl[\bigl(\hat{Y}_{\mathrm{ok}}(\bx_0) - Y(\bx_0)\bigr)^2\Bigr] = \operatorname{Var}\bigl(\hat{Y}_{\mathrm{ok}}(\bx_0) - Y(\bx_0)\bigr) + \underbrace{\Bigl(\mathbb{E}\bigl[\hat{Y}_{\mathrm{ok}}(\bx_0) - Y(\bx_0)\bigr]\Bigr)^2}_{= 0}\,.
\end{equation}
The same relationship holds for the SK predictor. The optimal weight vectors, $\boldsymbol{\lambda}_{\mathrm{ok}}$ and $\boldsymbol{\lambda}_{\mathrm{sk}}$, are obtained by minimising the corresponding MSPE. The detailed algebraic derivations are omitted here; interested readers are referred to \cite{webster2007} for complete details.

For OK, this minimisation problem is subject to the unbiasedness constraint. Solving it using the method of Lagrange multipliers leads to the linear system
\begin{equation}
	\begin{bmatrix}
		\mathbf{\Gamma} & \mathbf{1} \\
		\mathbf{1}^{\top} & 0
	\end{bmatrix}
	\begin{bmatrix}
		\boldsymbol{\lambda}_{\mathrm{ok}} \\
		\psi
	\end{bmatrix}
	=
	\begin{bmatrix}
		\boldsymbol{\gamma}_0 \\
		1
	\end{bmatrix},
	\label{ok_system}
\end{equation}
where $\mathbf{\Gamma} = \bigl[\gamma(\bx_j, \bx_k)\bigr]_{j,k=1}^n$ is the $n \times n$ semivariogram matrix, $\boldsymbol{\gamma}_0 = [\gamma(\bx_1, \bx_0), \ldots, \gamma(\bx_n, \bx_0)]^{\top}$ is the vector of semivariogram values between the observations and the target location, and $\psi$ is the Lagrange multiplier introduced to enforce the unbiasedness constraint. From \eqref{ok_system}, the weight vector $\boldsymbol{\lambda}_{\mathrm{ok}}$ can be expressed as
\begin{equation}
	\boldsymbol{\lambda}_{\mathrm{ok}} = \mathbf{\Gamma}^{-1} \bigl(\boldsymbol{\gamma}_0 - \mathbf{1}\psi \bigr),
\end{equation}
provided that $\mathbf{\Gamma}$ is non-singular. Using these weights, the spatial prediction $\hat{y}_{\mathrm{ok}}(\bx_0)$, referred to as the Kriging mean, is computed as
\begin{equation}
	\hat{y}_{\mathrm{ok}}(\bx_0) = \boldsymbol{\lambda}_{\mathrm{ok}}^{\top}\by_{\mathrm{obs}}.
	\label{eq:ok_prediction}
\end{equation}
The corresponding minimised prediction error variance $\sigma^2_{\mathrm{ok}}(\bx_0)$, known as the Kriging variance, is given by
\begin{equation}
	\sigma^2_{\mathrm{ok}}(\bx_0) = \boldsymbol{\lambda}_{\mathrm{ok}}^{\top}\boldsymbol{\gamma}_0 + \psi.
	\label{eq:ok_variance}
\end{equation}
Computing both $\hat{y}_{\mathrm{ok}}(\bx_0)$ and $\sigma^2_{\mathrm{ok}}(\bx_0)$ requires evaluating $\psi$; its analytical expression is derived in \Cref{app:deriving_lagrange}.

For SK, minimising the MSPE requires neither an unbiasedness constraint nor a Lagrange multiplier. Assuming that $\mathbf{\Gamma}$ is non-singular, the optimal weight vector is $\boldsymbol{\lambda}_{\mathrm{sk}} = \mathbf{\Gamma}^{-1}\boldsymbol{\gamma}_0$. The resulting SK prediction $\hat{y}_{\mathrm{sk}}(\bx_0)$ and Kriging variance $\sigma^2_{\mathrm{sk}}(\bx_0)$ are given by
\begin{align}
	\hat{y}_{\mathrm{sk}}(\bx_0) &= \mu_0 + \boldsymbol{\lambda}_{\mathrm{sk}}^{\top}(\by_{\mathrm{obs}} - \mu_0 \mathbf{1})\,, \label{eq:sk_prediction} \\
	\sigma^2_{\mathrm{sk}}(\bx_0) &= \boldsymbol{\lambda}_{\mathrm{sk}}^{\top}\boldsymbol{\gamma}_0. \label{eq:sk_variance}
\end{align}
%===================================================================================================
\section{Gaussian process regression}
\label{sec:GP_regression}
\begin{figure}[htbp] 
	\centering
	\includegraphics[width=0.49\textwidth]{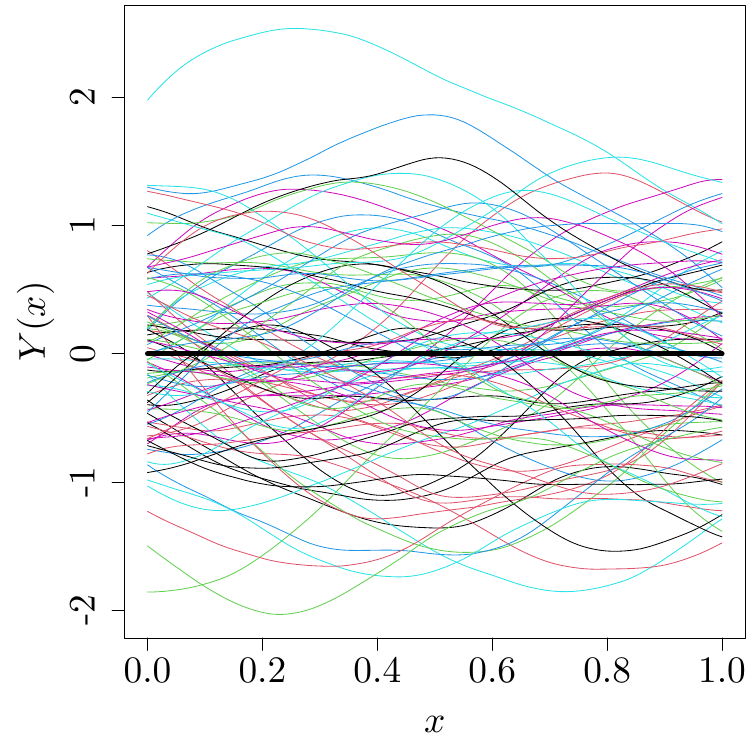} 
	\includegraphics[width=0.49\textwidth]{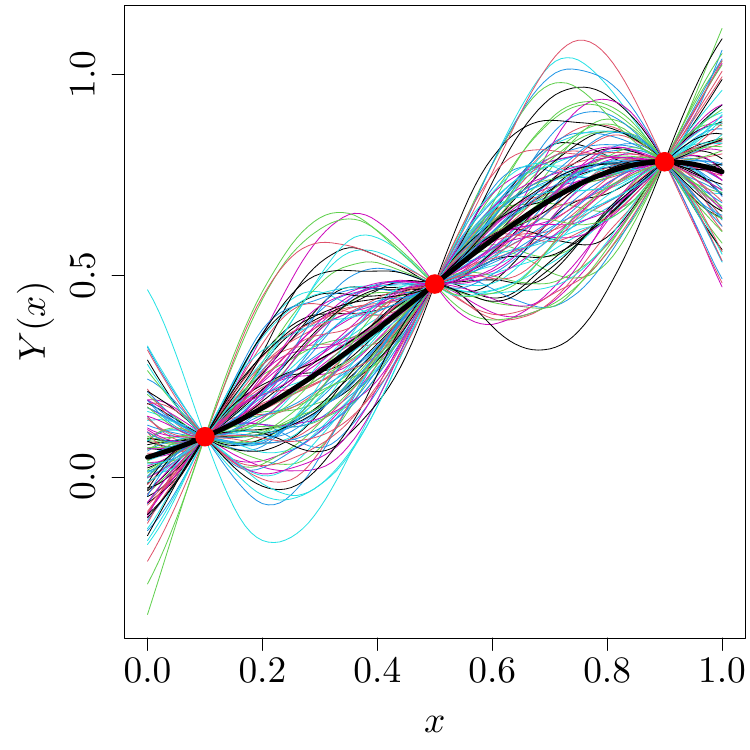} 
	\caption{Illustration of the prior (left) and posterior (right) GPs. The thin coloured lines are sample paths, while the thick black lines represent the process means. The posterior GP is conditioned on the observations (red dots).}
	\label{fig:GP_conditioning}
\end{figure}

%===================================================================================================
We now turn to GP regression, introducing it from a machine learning perspective and highlighting its differences from Kriging. GPs define a distribution over functions in a continuous domain, where any finite set of function values follows a multivariate Gaussian distribution \cite{GPML}. Within this framework, the underlying physical system is conceptualised as a particular realisation drawn from this prior process. Predictions are then obtained by conditioning the prior on available data. The resulting posterior process remains a GP, whose mean and variance can be calculated in closed form. This is illustrated in \Cref{fig:GP_conditioning}, which shows the prior (left) and posterior (right) GPs.

This probabilistic description of the unknown function yields a full posterior predictive distribution that explicitly quantifies epistemic uncertainty. This distinguishes GPs from the classical Kriging formulation, where prediction and the associated error variance are derived through the BLUP criterion without requiring distributional assumptions. Furthermore, unlike the traditional geographic setting of Kriging, the input domain $\mathcal{X}$ of a GP may represent a general, potentially high-dimensional space. A GP can therefore be viewed as a GRF indexed by a general input space $\mathcal{X}$, consistent with RF theory \cite{adler2007}.

A GP is fully specified by its mean and covariance functions. For a second-order stationary RF $Y(\bx)$ modelled as a GP, we write $Y(\bx) \sim \mathcal{GP}\bigl(\mu_0, C\bigr)$, where the covariance function $C$ depends only on the separation lag $\bh$. The predictive distribution at the target location $\bx_0$ is derived from the joint distribution of the random vector $\by$ and the unobserved value $Y(\bx_0)$: 
\begin{equation}
	\begin{bmatrix}
		\by \\
		Y(\bx_0)
	\end{bmatrix}
	\sim
	\mathcal{N}\left(
	\begin{bmatrix}
		\mu_0\mathbf{1} \\
		\mu_0
	\end{bmatrix},
	\begin{bmatrix}
		\bC & \bc(\bx_0) \\
		\bc(\bx_0)^\top & C(\bx_0, \bx_0)
	\end{bmatrix}
	\right),
\end{equation}
where $\bc(\bx_0) = \bigl[C(\bx_1, \bx_0), \dots, C(\bx_n, \bx_0)\bigr]^\top \in \mathbb{R}^n$ and $\bC = \bigl[C(\bx_j, \bx_k)\bigr]_{j,k=1}^n \in \mathbb{R}^{n \times n}$. By the properties of multivariate Gaussian distributions, conditioning on the observed data $\by_{\mathrm{obs}}$ yields
\begin{equation}
	Y(\bx_0) \mid \by = \by_{\mathrm{obs}} \sim \mathcal{N}\left(m_{\mathrm{sk}}(\bx_0), s_{\mathrm{sk}}^2(\bx_0)\right).
\end{equation}
where, for known $\mu_0$, the posterior mean and variance are
\begin{align}
	m_{\mathrm{sk}}(\bx_0) &= \mu_0 + \bc(\bx_0)^\top\bC^{-1}(\by_{\mathrm{obs}} - \mu_0\mathbf{1}), \label{eq:GP_mean} \\
	s_{\mathrm{sk}}^2(\bx_0) &= C(\bx_0, \bx_0) - \bc(\bx_0)^\top\bC^{-1}\bc(\bx_0). \label{eq:GP_var}
\end{align}

In practice, $\mu_0$ is often unknown and must be estimated from the data. We estimate $\mu_0$ using maximum likelihood estimation (MLE), as detailed in \Cref{sec:param_estim_GPs}. The predictive mean is obtained by replacing $\mu_0$ with its estimator $\hat{\mu}_0$:
\begin{equation} 
	m_{\mathrm{ok}}(\bx_0) = \hat{\mu}_0 + \bc(\bx_0)^\top\bC^{-1}(\by_{\mathrm{obs}} - \hat{\mu}_0\mathbf{1}),
	\label{eq:GP_mean_unknown} 
\end{equation} 
which has the same form as $m_{\mathrm{sk}}(\bx_0)$. Accounting for the uncertainty associated with the unknown mean gives the predictive variance \cite{roustant2012}
\begin{equation} 
	s_{\mathrm{ok}}^2(\bx_0) = C(\bx_0,\bx_0) -\bc(\bx_0)^\top\bC^{-1}\bc(\bx_0) +\frac{\left(1-\mathbf1^\top\bC^{-1}\bc(\bx_0)\right)^2} {\mathbf1^\top\bC^{-1}\mathbf1}. \label{eq:GP_var_unknown} 
\end{equation} 
The additional term in $s_{\mathrm{ok}}^2(\bx_0)$ accounts for the uncertainty associated with $\mu_0$ and increases the predictive variance relative to $s_{\mathrm{sk}}^2(\bx_0)$.
%===================================================================================================
\section{Valid variogram and covariance functions} 
\label{sec:variogram_covariance_fun}
%===================================================================================================
Not all functions can serve as valid covariance or variogram models for quantifying spatial dependence. For a candidate function to be valid, it must satisfy specific properties, as detailed below.
\subsection{Validity criteria and Bochner’s theorem}
\label{sub:validity_criteria}
A mapping $\gamma : \mathcal{X} \times \mathcal{X} \to \mathbb{R}$ is a valid semivariogram if it is symmetric, $\gamma(\bx, \bx') = \gamma(\bx', \bx)$, and conditionally negative semi-definite (CNSD). The function $\gamma$ is CNSD if, for any $n \in \mathbb{N}$, any set of locations $\{\bx_1, \ldots, \bx_n\} \subset \mathcal{X}$, and any coefficient vector $\boldsymbol{\alpha}_\gamma \in \mathbb{R}^n$ satisfying $\mathbf{1}^\top \boldsymbol{\alpha}_\gamma = 0$, the following inequality holds:
\begin{equation}\label{eq:variogram_cnsd}
	\boldsymbol{\alpha}_\gamma^\top \boldsymbol{\Gamma} \boldsymbol{\alpha}_\gamma \le 0.
\end{equation}

Similarly, a mapping $C : \mathcal{X} \times \mathcal{X} \to \mathbb{R}$ is a valid covariance function if it is symmetric, $C(\bx, \bx') = C(\bx', \bx)$, and positive semi-definite (PSD). The function $C$ is PSD if, for any $n \in \mathbb{N}$, any set of locations $\{\bx_1, \ldots, \bx_n\} \subset \mathcal{X}$, and any coefficient vector $\boldsymbol{\alpha}_c \in \mathbb{R}^n$, the following inequality holds:
\begin{equation}\label{eq:cov_psd}
	\boldsymbol{\alpha}_c^\top \mathbf{C} \boldsymbol{\alpha}_c \ge 0.
\end{equation}

If the inequalities in \eqref{eq:variogram_cnsd} and \eqref{eq:cov_psd} hold strictly for distinct locations and non-zero vectors ($\boldsymbol{\alpha}_\gamma \ne \mathbf{0}$ with $\mathbf{1}^\top \boldsymbol{\alpha}_\gamma = 0$, and $\boldsymbol{\alpha}_c \ne \mathbf{0}$), $\gamma$ and $C$ are \emph{conditionally negative definite} (CND) and \emph{positive definite} (PD), respectively. Strict positive definiteness of $C$ ensures that the associated covariance matrix is non-singular, which is important for the linear systems arising in GP regression.

Verifying positive semi-definiteness directly from the definition \eqref{eq:cov_psd} can be tedious. Bochner’s theorem \cite{bochner2016} provides a practical means of constructing stationary covariance functions through their spectral representation. More specifically, when a spectral density exists, a continuous function $C(\bh)$ is the covariance function of a second-order stationary RF if and only if it can be represented as
\begin{equation}
	C(\bh) = \frac{1}{(2\pi)^d} \int_{\mathbb{R}^d} e^{i \boldsymbol{\xi}^\top \bh}\, S(\boldsymbol{\xi})\, \mathrm{d}\boldsymbol{\xi}.
	\label{eq:bochner_theorem}
\end{equation}
Here, $S(\boldsymbol{\xi}) \ge 0$ is the spectral density and $\boldsymbol{\xi} \in \mathbb{R}^d$ denotes the spatial frequency vector. Under this representation, $C(\bh)$ and $S(\boldsymbol{\xi})$ form a Fourier transform pair. Consequently, a stationary covariance function can be validated by verifying that its corresponding spectral density is non-negative for all $\boldsymbol{\xi}\in\mathbb R^d$.

\begin{figure}[htbp] 
	\centering
	\includegraphics[width=0.49\textwidth]{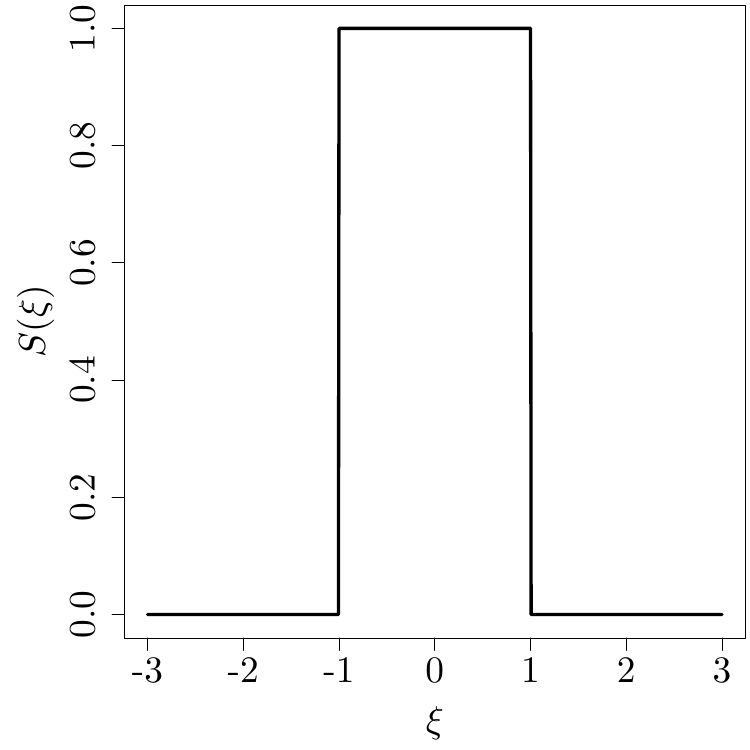} 
	\includegraphics[width=0.49\textwidth]{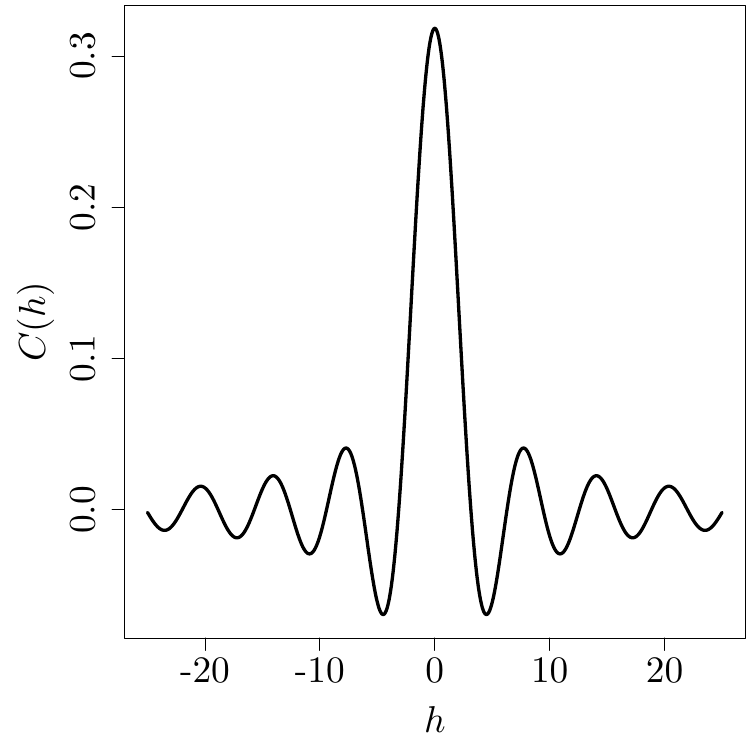} 
	\caption{Illustration of Bochner's theorem in one dimension. Spectral density $S(\xi)$ (left) and the corresponding covariance function $C(h)$ (right) obtained by Fourier transformation. The non-negativity of $S(\xi)$ ensures that $C(h)$ is a valid covariance function.}
	\label{fig:bochner}
\end{figure}

To illustrate Bochner's theorem, a one-dimensional example is provided in \Cref{fig:bochner}. The left plot shows the spectral density, which is constant on $[-1, 1]$:
\begin{equation}
	S(\xi) = 
	\begin{cases} 
		1 & \text{if } |\xi| \le 1, \\ 
		0 & \text{if } |\xi| > 1. 
	\end{cases}
\end{equation}
The corresponding covariance function (right plot) is obtained from \eqref{eq:bochner_theorem}:
\begin{equation}
	C(h) = \frac{1}{2\pi} \int_{-1}^{1} e^{i \xi h} \, \mathrm{d}\xi = \frac{\sin(h)}{\pi h}.
\end{equation}
Since $S(\xi) \ge 0$ for all $\xi \in \mathbb{R}$, $C(h)$ is guaranteed to be a valid covariance function.

\subsection{Common models for spatial dependence}
\label{sub:common_models}
Having established the validity criteria, we now focus on specific parametric models that are widely used for spatial modelling and inference. The analytical expressions for the most common isotropic semivariogram models are provided in \Cref{tab:variog_functions}. The corresponding covariance functions can be derived directly using the identity $\gamma(\bh) = C(\mathbf{0}) - C(\bh)$ (see \Cref{variog_cov_relation}). 

In machine learning, the parameters $\sigma_0^2$ and $\phi$ represent the \emph{process variance} and \emph{length-scale}, respectively. The former governs the amplitude of the process and the latter controls the characteristic scale of spatial dependence. In the absence of noise (see \Cref{sec:noisy_RF}), $\sigma_0^2$ is referred to as the \emph{sill} in geostatistical terminology. For the spherical model, $\phi$ also corresponds to the \emph{range} in geostatistics, as the semivariogram reaches the sill at $\|\bh\|=\phi$. For models that approach the sill asymptotically, a \emph{practical range} can instead be defined as the distance at which the semivariogram reaches a specified percentage (e.g., 95\%) of the sill.

\begin{table}[htbp]
	\centering
	\caption{Common isotropic semivariogram functions.}
	\label{tab:variog_functions}
	\begin{tabular}{ll}
		\toprule
		\textbf{Model Type} & \textbf{Semivariogram $\gamma(\|\bh\|)$} \\
		\midrule
		Spherical & $\begin{cases} 
			\sigma_0^2 \left[ \frac{3}{2}\left(\frac{\|\bh\|}{\phi}\right) - \frac{1}{2}\left(\frac{\|\bh\|}{\phi}\right)^3 \right], & \|\bh\| \le \phi \\
			\sigma_0^2, & \|\bh\| > \phi 
		\end{cases}$ \\
		\addlinespace[0.3cm]
		Exponential & $\sigma_0^2 \left[1 - \exp\left(-\frac{\|\bh\|}{\phi}\right)\right]$ \\
		\addlinespace[0.3cm]
		Gaussian & $\sigma_0^2 \left[1 - \exp\left(-\frac{\|\bh\|^2}{2\phi^2}\right)\right]$ \\
		\addlinespace[0.3cm]
		Matérn $3/2$ & $\sigma_0^2 \left[1 - \left(1 + \frac{\sqrt{3}\|\bh\|}{\phi}\right) \exp\left(-\frac{\sqrt{3}\|\bh\|}{\phi}\right)\right]$ \\
		\addlinespace[0.3cm]
		Matérn $5/2$ & $\sigma_0^2 \left[1 - \left(1 + \frac{\sqrt{5}\|\bh\|}{\phi} + \frac{5\|\bh\|^2}{3\phi^2}\right) \exp\left(-\frac{\sqrt{5}\|\bh\|}{\phi}\right)\right]$ \\
		\bottomrule
	\end{tabular}
\end{table}

Most of the spatial dependence models in \Cref{tab:variog_functions} can be derived from the Matérn kernel. In its isotropic form, the Matérn covariance is defined as
\begin{equation}\label{eq:matern_cov}
	C\bigl(\|\bh\|\bigr) = \sigma_0^2 \frac{2^{1-\nu}}{\Gamma(\nu)} \left(\sqrt{2\nu}\,\frac{\|\bh\|}{\phi}\right)^{\nu} K_{\nu}\left(\sqrt{2\nu}\,\frac{\|\bh\|}{\phi}\right),
\end{equation}
where $\Gamma(\cdot)$ is the gamma function and $K_\nu(\cdot)$ is the modified Bessel function of the second kind of order $\nu$. When $\nu$ takes half-integer values, $K_\nu(\cdot)$ reduces to a finite sum of elementary functions, and the resulting Matérn covariance can be expressed as the product of a polynomial and an exponential (e.g., the Matérn 3/2 and 5/2 semivariogram models in \Cref{tab:variog_functions}). Furthermore, the exponential model corresponds to the Matérn covariance for $\nu = 1/2$, while the Gaussian model is obtained as the limiting case $\nu \to \infty$.

\begin{figure}[htbp]
	\centering
	\includegraphics[width=0.49\textwidth]{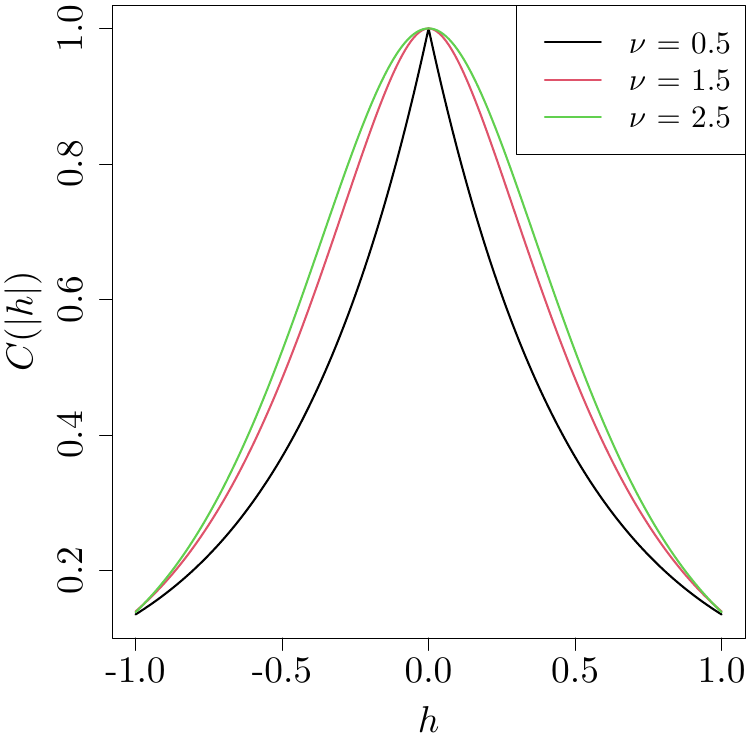}
	\includegraphics[width=0.49\textwidth]{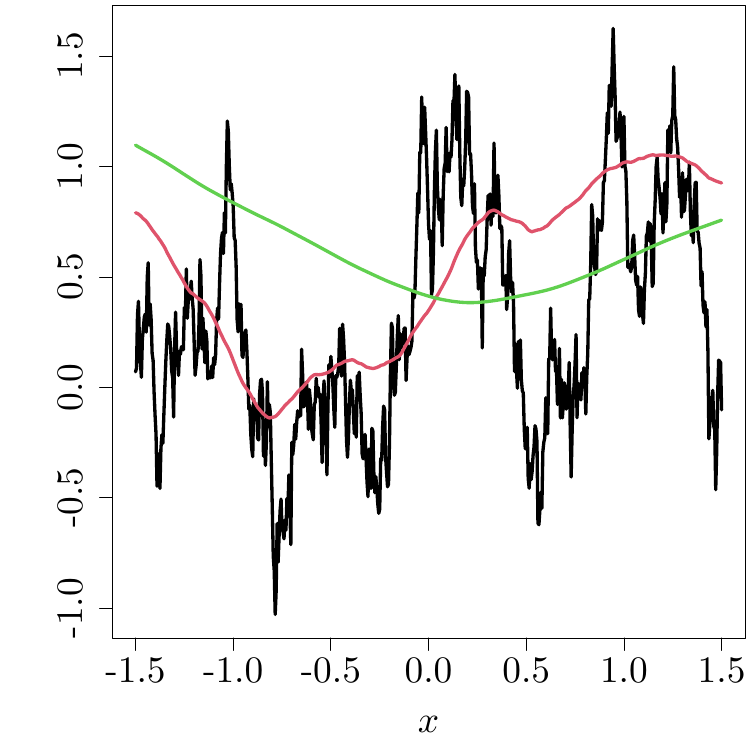}
	\caption{Left: Matérn covariance functions for three values of the parameter $\nu$ (0.5, 1.5, and 2.5). Right: sample paths associated with $\nu = 0.5$ (black), $\nu = 1.5$ (red), and $\nu = 2.5$ (green). Increasing $\nu$ produces smoother covariance functions near the origin and smoother sample paths. Here, $\sigma_0^2 = 1$ and $\phi = 0.5$.}
	\label{fig:matern_plots}
\end{figure}

The parameter $\nu > 0$ controls the smoothness of the spatial process, such that an RF with this covariance is $r$ times MS differentiable for all integers $r < \nu$. \Cref{fig:matern_plots} illustrates the impact of $\nu$ on both the Matérn covariance function and its associated process realisations for $\sigma_0^2 = 1$ and $\phi = 0.5$. On the left, the Matérn kernel is plotted for three values of $\nu$ (0.5, 1.5, and 2.5). Increasing $\nu$ makes $C(h)$ smoother near $h = 0$, changing its behaviour at the origin from a sharp cusp to a flat curve. On the right, this increased smoothness of the covariance function is reflected in the sample paths, which become smoother as $\nu$ increases.

%===================================================================================================
\section{Noisy observations and the nugget effect}
\label{sec:noisy_RF}
%===================================================================================================
Real-world data are often contaminated with noise, representing erratic, spatially uncorrelated fluctuations. Such fluctuations cannot be explained by the spatial dependence structure of the underlying RF. The \emph{nugget effect} is commonly used to account for this additional variability. A simple representation of the noisy observation process is given by the RF $\tilde{Y}(\bx)$:
\begin{equation}
	\tilde{Y}(\bx) = Y(\bx) + W(\bx),
	\label{eq:noisy_rf}
\end{equation}
in which $W(\bx)$ is a zero-mean, spatially uncorrelated process that is assumed to be independent of $Y(\bx)$. Let $\tau^2 > 0$ denote the variance of $W(\bx)$, known as the nugget variance. In this work, we assume $\tau^2$ is constant across the domain. The heteroscedastic case, where $\tau^2$ varies with $\bx$, is discussed in \cite{binois2018}.

The nugget effect can be attributed to pure measurement error (or observation noise), arising from instrument inaccuracy and operational errors. It can also capture microscale spatial variation occurring at distances smaller than the sampling resolution. We denote these components by $\epsilon(\bx)$ and $\eta(\bx)$, respectively. When both components are present, $W(\bx)$ can be decomposed as $W(\bx) = \epsilon(\bx) + \eta(\bx)$. Note that some literature reserves the term \emph{nugget effect} strictly for the microscale component $\eta(\bx)$ and does not consider it noise. For example, the \textsf{DiceKriging} package \cite{roustant2012} in R provides two separate arguments: \texttt{nugget}, to model fine-scale variability, and \texttt{noise.var}, to account for observation noise.

The nugget effect changes the behaviour of both the covariance and semivariogram functions by introducing a discontinuity at the origin. To illustrate this, let $\tilde{C}(\bh)$ and $\tilde{\gamma}(\bh)$ denote the covariance function and semivariogram of $\tilde{Y}(\bx)$, respectively. Since $W(\bx)$ is spatially uncorrelated, these functions take the form:
\begin{align}
	\tilde{C}(\bh) &= C(\bh) + \tau^2 \delta_{\bh, \mathbf{0}}, \label{eq:noisy_cov} \\
	\tilde{\gamma}(\bh) &= \gamma(\bh) + \tau^2 (1 - \delta_{\bh, \mathbf{0}}), \label{eq:noisy_variog}
\end{align}
where $\delta_{\bh, \mathbf{0}}$ denotes the Kronecker delta, with $\delta_{\bh, \mathbf{0}} = 1$ when $\bh = \mathbf{0}$ and $0$ otherwise.

In \eqref{eq:noisy_cov}, the total variance at zero lag is $\tilde{C}(\mathbf{0}) = \sigma_0^2 + \tau^2$, while the spatial covariance structure remains completely unchanged for any non-zero $\bh$. Specifically,
\[
\lim_{\|\bh\|\to 0^+} \tilde{C}(\bh) = C(\mathbf{0}) = \sigma_0^2, 
\]
which demonstrates the discontinuity at the origin. A similar behaviour is observed for the semivariogram. According to \eqref{eq:noisy_variog}, the nugget effect shifts the semivariogram upwards by $\tau^2$ for all $\bh \neq \mathbf{0}$. Therefore,
\[
\lim_{\|\bh\|\to 0^+} \tilde{\gamma}(\bh) = \tau^2,
\]
while $\tilde{\gamma}(\mathbf{0}) = 0$ by definition. In this setting, the sill is $\tau^2 + \sigma_0^2$, with $\sigma_0^2$ referred to as the \emph{partial sill}.

We next examine how the nugget effect modifies the predictive expressions of Kriging and GP regression. Our focus is on predicting the latent process $Y(\bx_0)$ rather than the noisy process $\tilde{Y}(\bx_0)$. In both approaches, prediction is based on the noisy observations
\begin{equation}
	 \tilde{\by}_{\mathrm{obs}} = \bigl[\tilde{y}(\bx_1), \ldots, \tilde{y}(\bx_n)\bigr]^\top.
\end{equation}

In the Kriging equations, the nugget effect adjusts the semivariogram matrix as follows: 
\begin{equation} 
	\tilde{\mathbf{\Gamma}} = \mathbf{\Gamma} + \tau^2(\mathbf{1}\mathbf{1}^{\top}-\mathbf{I}),
\end{equation} 
where $\mathbf{I}$ is the $n \times n$ identity matrix. The diagonal elements remain zero by definition, whereas each off-diagonal element is increased by $\tau^2$. This is consistent with the effect of the nugget on the semivariogram function, which exhibits a discontinuity at the origin. Since the prediction target is the latent process $Y(\bx_0)$, the vector $\boldsymbol{\gamma}_0$, containing semivariogram values of the latent process between the sampled locations and $\bx_0$, remains unchanged.

The nugget effect modifies the covariance structure of GPs differently. It adds $\tau^2$ to the diagonal elements of the training covariance matrix, while the off-diagonal elements remain unaffected: 
\begin{equation} 
	\tilde{\mathbf{C}} = \mathbf{C} + \tau^2 \mathbf{I}. 
\end{equation} 
Since the prediction target is the latent process $Y(\bx_0)$, the cross-covariance vector $\bc(\bx_0)$ does not change. Thus, the nugget affects the covariance structure among the noisy observations, while their covariance with the latent prediction target remains the same.

From a computational perspective, the nugget effect improves the numerical stability of computations involving the training covariance matrix. For example, the use of a small nugget has been advocated even for deterministic computer experiments \cite{gramacy2012}. When the sampled locations are clustered, or when a very smooth covariance function such as the squared exponential is used, $\bC$ can become ill-conditioned. The condition number of a PD covariance matrix $\bC$ is defined as
\[
\operatorname{cond}(\bC) = \frac{\eta_\textsf{max}}{\eta_\textsf{min}},
\]
where $\eta_\textsf{max}$ and $\eta_\textsf{min}$ are its largest and smallest eigenvalues, respectively. The diagonal adjustment $\tilde{\bC} = \bC + \tau^2\mathbf{I}$ shifts all eigenvalues of $\bC$ by $\tau^2$, including $\eta_\textsf{max}$ and $\eta_\textsf{min}$. This reduces the condition number of $\tilde{\bC}$ and improves its numerical stability. Such an adjustment is commonly referred to as \emph{regularisation}; see \cite{mohammadi2017} for further details.
%===================================================================================================
\section{Estimating unknown parameters}
\label{sec:param_estim}
%===================================================================================================
In practice, the parameters of the spatial models introduced in the previous sections are unknown and must be estimated from observed data. This section outlines two common estimation paradigms, namely least squares estimation (LSE) and MLE. Typically, the Kriging parameters are estimated via LSE, while MLE is used within the GP framework. For simplicity, we assume isotropy throughout this section.
%===================================================================================================
\subsection{LSE for Kriging}
\label{sec:param_estim_kriging}
%===================================================================================================
LSE is a classical geostatistical approach, which fits a parametric semivariogram model (e.g., those in \Cref{tab:variog_functions}) to the empirical semivariogram. Let $\boldsymbol{\theta}=(\sigma_0^2, \phi, \tau^2)^\top$ be the vector of unknown parameters and $\hat{\boldsymbol{\theta}}$ denote its estimate. In the LSE method, $\hat{\boldsymbol{\theta}}$ is obtained by minimising a loss function measuring the discrepancy between the theoretical semivariogram and its empirical estimates. 

The empirical semivariogram is constructed from pairwise semivariances. For every pair of sampled locations $\bx_j$ and $\bx_k$ with corresponding observed values $\tilde{y}(\bx_j)$ and $\tilde{y}(\bx_k)$, the semivariance is computed as
\begin{equation}
	v_{jk} = \frac{1}{2}\left(\tilde{y}(\bx_j)-\tilde{y}(\bx_k)\right)^2, \qquad 1 \le j < k \le n.
\end{equation}
Since $v_{jk}$ is based on differences between observations, the constant mean $\mu_0$ is eliminated and therefore does not need to be estimated in the LSE procedure.

The $v_{jk}$ values are then plotted against their corresponding lag distances $\|\bx_j-\bx_k\|$, yielding the \emph{variogram cloud}, which is often noisy and difficult to interpret. To address this issue, pairs of locations with similar lag distances are grouped into spatial distance bins, and the average semivariance within each bin defines the empirical semivariogram.

More specifically, suppose $L$ spatial distance bins $B_1, B_2, \ldots, B_L$ are constructed, which may be of equal or varying widths. The set of pair indices falling within the $l$th bin $B_l$ is defined as:
\begin{equation}
	N_l = \bigl\{(j,k) : 1 \le j < k \le n, \ \|\bx_j-\bx_k\| \in B_l\bigr\}.
\end{equation}
The empirical semivariogram estimator based on Matheron's method of moments \cite{cressie1993} is given by:
\begin{equation}
	\hat{\gamma}_l = \frac{1}{|N_l|}\sum_{(j,k)\in N_l}v_{jk}, \qquad l=1,\ldots,L,
\end{equation}
with $|N_l|$ being the cardinality of $N_l$.

Given a parametric isotropic semivariogram model $\gamma(\|\bh\|;\boldsymbol{\theta})$, the parameter vector $\boldsymbol{\theta}$ can be estimated using weighted LSE:
\begin{equation}
	\hat{\boldsymbol{\theta}} = \arg\min_{\boldsymbol{\theta}} \sum_{l=1}^{L}w_l\Bigl(\hat{\gamma}_l - \gamma(\bar{h}_l;\boldsymbol{\theta})\Bigr)^2,
\end{equation}
where $w_l$ are weighting factors, often chosen to be proportional to $|N_l|$ \cite{cressie1993}, and $\bar{h}_l$ is the average lag distance within bin $B_l$, given by:
\begin{equation}
	\bar{h}_l = \frac{1}{|N_l|}\sum_{(j,k)\in N_l} \|\bx_j - \bx_k\|.
\end{equation}
Setting $w_l = 1$ for all $l$ simplifies the objective function to ordinary LSE. Once $\hat{\boldsymbol{\theta}}$ is obtained, the fitted semivariogram $\gamma(\|\bh\|;\hat{\boldsymbol{\theta}})$ is used to form the Kriging system.

\begin{figure}[htbp]
	\centering
	\includegraphics[width=0.49\textwidth]{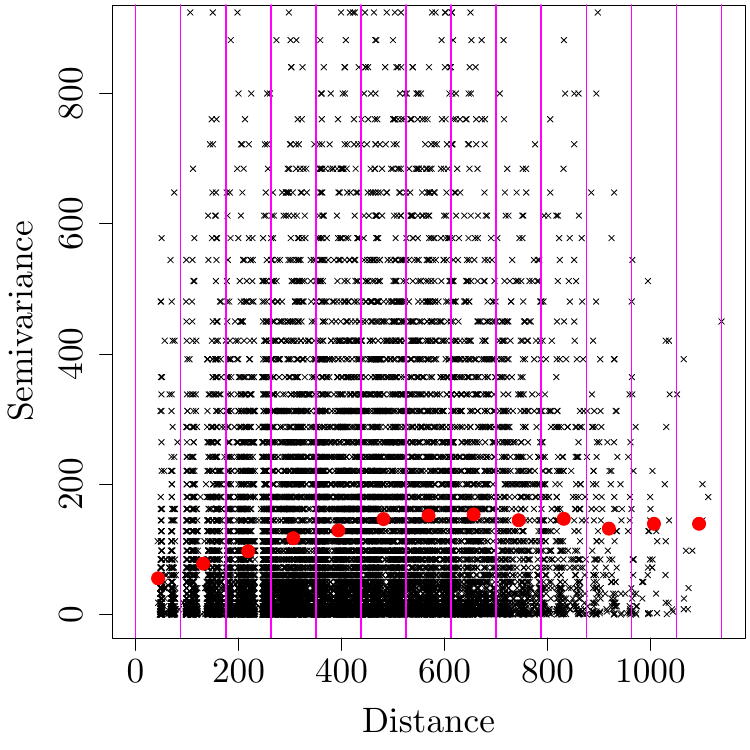}
	\includegraphics[width=0.49\textwidth]{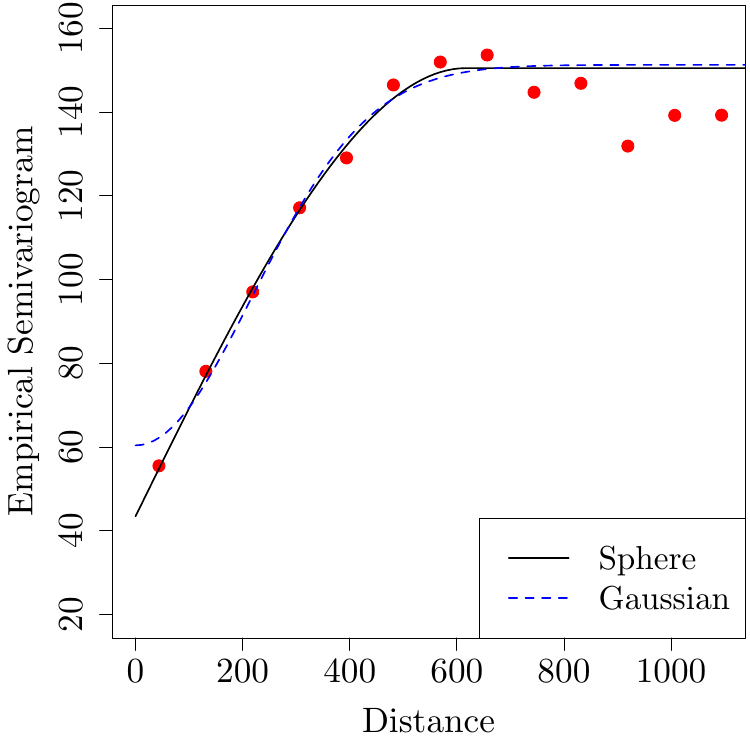}
	\caption{Illustration of the LSE procedure for semivariogram fitting. Left: variogram cloud for the calcium dataset. Right: empirical semivariogram (red dots) and fitted spherical (black) and Gaussian (blue dashed) semivariogram models.}
	\label{fig:LSE}
\end{figure}

The LSE procedure is illustrated in \Cref{fig:LSE} using a spatial dataset comprising calcium concentrations measured in soil samples, available in the \textsf{geoR} package \cite{geoR}. The left panel shows the variogram cloud computed by the \texttt{variog} function in \textsf{geoR}. The right panel presents the fits of two theoretical semivariogram models, namely the spherical (black) and Gaussian (blue dashed) models, to the empirical semivariogram (red dots). The models are fitted with the \texttt{variofit} function.
%===================================================================================================
\subsection{MLE for GPs}
\label{sec:param_estim_GPs}
%===================================================================================================
MLE is one of the most popular parameter estimation techniques in statistics due to its desirable asymptotic properties, including consistency and efficiency under suitable regularity conditions \cite{myung2003}. It is also widely implemented in software packages such as \textsf{DiceKriging} in R \cite{roustant2012} and \textsf{GPflow} in Python \cite{matthews2017}. We next introduce MLE for estimating $\boldsymbol{\theta}$ together with the constant GP mean $\mu_0$.

Unlike LSE, which does not require specification of a full probability distribution, MLE is based on the joint distribution of the observations. Under the GP model, the observations are regarded as a realisation of a multivariate Gaussian random vector with mean $\mu_0\mathbf{1}$ and covariance matrix $\tilde{\bC}$, which depends on $\boldsymbol{\theta}$. The corresponding joint probability density function, evaluated at the observed vector $\tilde{\by}_{\mathrm{obs}}$, defines the likelihood function. For computational convenience, we work with the log-likelihood:
\begin{equation}
	\log \mathcal{L}(\mu_0, \boldsymbol{\theta}) = - \frac{n}{2} \log(2\pi) - \frac{1}{2}\log \det(\tilde{\bC}) -\frac{1}{2} \bigl(\tilde{\by}_{\mathrm{obs}} - \mu_0\mathbf{1}\bigr)^\top \tilde{\bC}^{-1} \bigl(\tilde{\by}_{\mathrm{obs}} - \mu_0\mathbf{1}\bigr).
	\label{eq:log_like}
\end{equation}

To obtain $\hat{\mu}_0$, we differentiate $\log \mathcal{L}(\mu_0, \boldsymbol{\theta})$ with respect to $\mu_0$ and set the resulting derivative to zero. This leads to the closed-form estimator:
\begin{equation}
	\hat{\mu}_0(\boldsymbol{\theta}) = \frac{\mathbf{1}^\top \tilde{\bC}^{-1} \tilde{\by}_{\mathrm{obs}}}{\mathbf{1}^\top \tilde{\bC}^{-1} \mathbf{1}}. \label{eq:mle_mu}
\end{equation}
This expression is then substituted for $\mu_0$ in \eqref{eq:log_like}, reducing the optimisation problem to $\boldsymbol{\theta}$. After omitting additive constants independent of $\boldsymbol{\theta}$, we obtain the concentrated (or profile) log-likelihood function. Since this function is nonlinear in $\boldsymbol{\theta}$, an analytical solution is not generally available, and numerical optimisation is typically required. The resulting likelihood surface can be multimodal, making it challenging to find its global maximum. Various optimisation algorithms have been proposed to address this problem; see, e.g., \cite{toal2011, butler2014}. 

A major limitation of MLE is its computational burden. At each iteration of the optimisation, a set of candidate parameter values is considered, from which the standard dense GP covariance matrix $\tilde{\bC}$ is constructed and inverted. Since $\tilde{\bC}$ changes with $\boldsymbol{\theta}$, these matrix operations are repeated as the optimisation proceeds. The computational and memory complexities associated with these operations are $\mathcal{O}(n^3)$ and $\mathcal{O}(n^2)$, respectively. Consequently, MLE can become prohibitively expensive as the sample size increases. To alleviate these costs, various approaches have been developed, including inducing-point methods \cite{titsias2009} and Vecchia approximations \cite{katzfuss2021}.
%===================================================================================================
\section{Conclusion}
\label{sec:conclusion}
%===================================================================================================
In this work, we explore geostatistical and machine learning perspectives on spatial modelling and prediction. Kriging and GP regression are presented through a unified mathematical framework grounded in random field theory. Our comparison reveals that the key differences between the two approaches lie in their underlying assumptions and practical implementations. Kriging relies on the semivariogram to quantify spatial dissimilarity without requiring specification of a full probability distribution. The unknown parameters are estimated using weighted least squares, yielding interpretable estimates of quantities such as the sill, range, and nugget. In contrast, a GP is fully specified by its mean and covariance functions. It can be viewed as a generalisation of the multivariate Gaussian distribution to function space, with the spatial process regarded as a particular realisation. The GP parameters are commonly estimated via maximum likelihood, which is based on the joint distribution of the observations. 

By presenting Kriging and GP regression within a common mathematical framework, this work clarifies how concepts developed in geostatistics and machine learning relate to one another and can be interpreted across the two disciplines.  This connection enables practitioners to benefit from both the interpretability of geostatistical methods and methodological developments in GPs, including advances in uncertainty quantification. Bringing these perspectives together facilitates the transfer of concepts, methods, and computational tools between the two fields. It also provides a more comprehensive foundation for spatial modelling and inference.
%===================================================================================================
\appendix

\section{Derivation of the Lagrange multiplier}
\label{app:deriving_lagrange}
Expanding the matrix system in \eqref{ok_system} yields the following two simultaneous equations:
\begin{align}
	\label{eq:row1}
	\mathbf{\Gamma}\boldsymbol{\lambda}_{\mathrm{ok}} + \psi\mathbf{1}
	&= \boldsymbol{\gamma}_0, \\
	\mathbf{1}^{\top}\boldsymbol{\lambda}_{\mathrm{ok}} &= 1.
	\label{eq:row2}
\end{align}
To solve for $\psi$, we first isolate $\boldsymbol{\lambda}_{\mathrm{ok}}$ in \Cref{eq:row1}. To this end, the vector $\psi\mathbf{1}$ is subtracted from both sides of the equation, and the resulting equation is multiplied from the left by $\mathbf{\Gamma}^{-1}$:
\begin{equation}
	\boldsymbol{\lambda}_{\mathrm{ok}} = \mathbf{\Gamma}^{-1}\boldsymbol{\gamma}_0 
	- \mathbf{\Gamma}^{-1}\mathbf{1}\psi.
	\label{eq:lambda_isolated}
\end{equation}
This expression is substituted for $\boldsymbol{\lambda}_{\mathrm{ok}}$ in the unbiasedness constraint given by \Cref{eq:row2}:
\begin{equation}
	\mathbf{1}^{\top} \left( \mathbf{\Gamma}^{-1}\boldsymbol{\gamma}_0 
	- \mathbf{\Gamma}^{-1}\mathbf{1}\psi \right) = 1.
\end{equation}
Distributing the row vector $\mathbf{1}^{\top}$ across the terms on the left-hand side gives:
\begin{equation}
	\mathbf{1}^{\top}\mathbf{\Gamma}^{-1}\boldsymbol{\gamma}_0 
	- \mathbf{1}^{\top}\mathbf{\Gamma}^{-1}\mathbf{1}\psi = 1.
\end{equation}
We then isolate the term containing $\psi$ by moving the first term to the right-hand side:
\begin{equation}
	-\mathbf{1}^{\top}\mathbf{\Gamma}^{-1}\mathbf{1}\psi 
	= 1 - \mathbf{1}^{\top}\mathbf{\Gamma}^{-1}\boldsymbol{\gamma}_0.
\end{equation}
Finally, the analytical expression for the Lagrange multiplier is obtained by dividing both sides by $-\mathbf{1}^{\top}\mathbf{\Gamma}^{-1}\mathbf{1}$:
\begin{equation}
	\psi = \frac{\mathbf{1}^{\top}\mathbf{\Gamma}^{-1}\boldsymbol{\gamma}_0 
		- 1}{\mathbf{1}^{\top}\mathbf{\Gamma}^{-1}\mathbf{1}}.
\end{equation}
%===================================================================================================

\bibliography{biblio}

@Article{tobler1970,
	ISSN = {00130095, 19448287},
	author = {W. R. Tobler},
	journal = {Economic Geography},
	pages = {234--240},
	publisher = {[Clark University, Wiley]},
	title = {A Computer Movie Simulating Urban Growth in the {D}etroit Region},
	urldate = {2026-04-22},
	volume = {46},
	year = {1970}
}

@Book{adler2007,
	title={Random Fields and Geometry},
	author = {Adler, Robert J. and Taylor, Jonathan E.},
	year={2007},
	publisher={Springer},
	series = {Springer Monographs in Mathematics},
	isbn = {9780387481128}
}

@Article{matheron1973, 
	title={The intrinsic random functions and their applications},
	volume={5}, 
	DOI={10.2307/1425829}, 
	number={3}, 
	journal={Advances in Applied Probability}, 
	author={Matheron, G.}, 
	year={1973}, 
	pages={439–468}
}

@Book{stein1999,
	author = {Stein, Michael L.},
	doi = {10.1007/978-1-4612-1494-6},
	isbn = {0-387-98629-4},
	note = {Some Theory for Kriging},
	pages = {xviii+247},
	series = {Springer Series in Statistics},
	title = {Interpolation of spatial data},
	address = {New York},
	year = {1999}
}

@Book{GPML,
	author = {Rasmussen, Carl Edward and Williams, Christopher K. I.},
	title = {Gaussian processes for machine learning (adaptive computation and machine learning)},
	year = {2005},
	isbn = {026218253X},
	publisher = {The MIT Press},
}

@Book{santner2018,
	Author = {Thomas J. Santner and Brian J. Williams and William I. Notz},
	Title = {The Design and Analysis of Computer Experiments, Second Edition},
	Year = {2018},
	Publisher = {Springer-Verlag},
	Pages = {436}
}

@Book{gramacy2020,
	title = {Surrogates: {G}aussian Process Modeling, Design and \
	Optimization for the Applied Sciences},
	author = {Robert B. Gramacy},
	publisher = {Chapman Hall/CRC},
	address = {Boca Raton, Florida},
	note = {\url{http://bobby.gramacy.com/surrogates/}},
	year = {2020}
}

@Book{cressie1993,
	title={Statistics for spatial data},
	author={Cressie, N.A.C.},
	isbn={9780471002550},
	lccn={gb93056987},
	series={Wiley series in probability and mathematical statistics: Applied probability and statistics},
	year={1993},
	publisher={J. Wiley}
}

@Article{roustant2012,
	author = {Olivier Roustant and David Ginsbourger and Yves Deville},
	title = {{DiceKriging}, {DiceOptim}: two {R} packages for the analysis of computer experiments by kriging-based metamodeling and optimization},
	journal = {Journal of Statistical Software},
	volume = {51},
	number = {1},
	year = {2012},
	pages = {1--55},
	doi = {10.18637/jss.v051.i01},
}

@Article{krige1953,
  title="{A statistical approach to some basic mine valuation problems on the Witwatersrand}",
author={Krige, Daniel G},
journal={Journal of the Southern African Institute of Mining and Metallurgy},
volume={52},
number={6},
pages={119--139},
year={1951},
publisher={Southern African Institute of Mining and Metallurgy}
}

@misc{mohammadi2017,
	title="{An analytic comparison of regularization methods for Gaussian processes}", 
	author={Hossein Mohammadi and Rodolphe Le Riche and Nicolas Durrande and Eric Touboul and Xavier Bay},
	year={2017},
	eprint={1602.00853},
	archivePrefix={arXiv},
	primaryClass={math.OC}
}

@Book{bochner2016,
	title = {Lectures on Fourier Integrals},
	author = {Salomon Bochner},
	publisher = {Princeton University Press},
	address = {Princeton},
	doi = {doi:10.1515/9781400881994},
	isbn = {9781400881994},
	year = {2016}
}

@book{hristopulos2020,
	author = {Hristopulos, Dionissios},
	year = {2020},
	title = {Random Fields for Spatial Data Modeling: A Primer for Scientists and Engineers},
	isbn = {ISBN 978-94-024-1916-0},
	publisher={Springer},
	doi = {10.1007/978-94-024-1918-4}
}

@book{webster2007,
	title={Geostatistics for environmental scientists},
	author={Webster, Richard and Oliver, Margaret A},
	year={2007},
	publisher={John Wiley \& Sons}
}

@book{pyrcz2014,
	author    = {Pyrcz, Michael J. and Deutsch, Clayton V.},
	title     = {Geostatistical Reservoir Modeling},
	edition   = {2nd},
	publisher = {Oxford University Press},
	address   = {New York},
	year      = {2014},
	pages     = {448},
	isbn      = {978-0199731442}
}

@article{kim2025,
	author = {Kim, Soo-Yeon and Kerr, Gaige Hunter and van Donkelaar, Aaron and Martin, Randall V. and West, J. Jason and Anenberg, Susan C.},
	doi = {10.1038/s43247-025-02270-9},
	isbn = {2662-4435},
	journal = {Communications Earth \& Environment},
	number = {1},
	pages = {311},
	title = "{Tracking air pollution and CO2 emissions in 13,189 urban areas worldwide using large geospatial datasets}",
	volume = {6},
	year = {2025}
}

@article{roy2017,
	author = {Roy, P. S. and Behera, M. D. and Srivastav, S. K.},
	isbn = {2250-1762},
	journal = {Proceedings of the National Academy of Sciences, India Section A: Physical Sciences},
	number = {4},
	pages = {465--472},
	title = {Satellite Remote Sensing: Sensors, Applications and Techniques},
	volume = {87},
	year = {2017}
}

@article{sun2020,
	author = {Sun, Shiquan and Zhu, Jiaqiang and Zhou, Xiang},
	isbn = {1548-7105},
	journal = {Nature Methods},
	number = {2},
	pages = {193--200},
	title = {Statistical analysis of spatial expression patterns for spatially resolved transcriptomic studies},
	volume = {17},
	year = {2020}
}

@ARTICLE{shahriari2016,
	author={Shahriari, Bobak and Swersky, Kevin and Wang, Ziyu and Adams, Ryan P. and de Freitas, Nando},
	journal={Proceedings of the IEEE}, 
	title={Taking the Human Out of the Loop: A Review of {B}ayesian Optimization}, 
	year={2016},
	volume={104},
	number={1},
	pages={148-175}
}

@article{chen2021,
	title = "{Solving and learning nonlinear PDEs with Gaussian processes}",
	journal = {Journal of Computational Physics},
	volume = {447},
	pages = {110668},
	year = {2021},
	issn = {0021-9991},
	author = {Yifan Chen and Bamdad Hosseini and Houman Owhadi and Andrew M. Stuart}
}

@article{matheron1963,
	author = {Matheron, Georges},
	doi = {10.2113/gsecongeo.58.8.1246},
	journal = {Economic Geology},
	number = {8},
	pages = {1246--1266},
	publisher = {Society of Economic Geologists},
	title = {Principles of geostatistics},
	volume = {58},
	year = {1963}
}

@InProceedings{cosier2024,
	title = {A Unifying Variational Framework for {G}aussian Process Motion Planning},
	author =  {Cosier, Lucas C. and Iordan, Rares and Zwane, Sicelukwanda N. T. and Franzese, Giovanni and Wilson, James T. and Deisenroth, Marc and Terenin, Alexander and Bekiroglu, Yasemin},
	booktitle = {Proceedings of The 27th International Conference on Artificial Intelligence and Statistics},
	pages = {1315--1323},
	year = {2024},
	volume = {238},
	series = {Proceedings of Machine Learning Research},
	publisher =    {PMLR}
}

@article{haran2011,
	title={Gaussian random field models for spatial data},
	author={Haran, Murali},
	journal={Handbook of Markov Chain Monte Carlo},
	pages={449--478},
	year={2011},
	publisher={CRC Press Boca Raton, FL, USA}
}

@book{moraga2023,
	title     = {Spatial Statistics for Data Science: Theory and Practice with {R}},
	author    = {Moraga, Paula},
	year      = {2023},
	publisher = {Chapman and Hall/CRC},
	series    = {Chapman \& Hall/CRC Data Science Series},
	isbn      = {9781032633510}
}

@article{geoR,
	author = {Ribeiro, Jr., Paulo J. and Diggle, Peter J.},
	title = "{The R Journal: geoR: A package for geostatistical analysis}",
	journal = {R News},
	year = {2001},
	note = {https://journal.r-project.org/articles/RN-2001-013/},
	volume = {1},
	issue = {2},
	issn = {1609-3631},
	pages = {14-18}
}

@article{myung2003,
	title={Tutorial on maximum likelihood estimation},
	author={Myung, In Jae},
	journal={Journal of mathematical Psychology},
	volume={47},
	number={1},
	pages={90--100},
	year={2003},
	publisher={Elsevier}
}

@article{marinescu2024,
	title="{Explaining and connecting Kriging with Gaussian process regression}",
	author={Marinescu, Marius},
	journal={arXiv preprint arXiv:2408.02331},
	year={2024}
}

@inproceedings{suryasentana2023,
	title="{Demystifying the connections between Gaussian process regression and kriging}",
	author={Suryasentana, SK and Sheil, BB},
	booktitle={The 9th International Conference on Offshore Site Investigation and Geotechnics},
	publisher={Society for Underwater Technology},
	pages={220--227},
	year={2023}
}

@inproceedings{titsias2009,
	title="{Variational learning of inducing variables in sparse Gaussian processes}",
	author={Titsias, Michalis},
	booktitle={Artificial intelligence and statistics},
	pages={567--574},
	year={2009},
	organization={PMLR}
}

@article{katzfuss2021,
	title="{A general framework for Vecchia approximations of Gaussian processes}",
	author={Katzfuss, Matthias and Guinness, Joseph},
	journal={Statistical Science},
	volume={36},
	number={1},
	pages={124--141},
	year={2021},
	publisher={JSTOR}
}

@article{butler2014,
	title = "{Efficient optimization of the likelihood function in Gaussian process modelling}",
	journal = {Computational Statistics \& Data Analysis},
	volume = {73},
	pages = {40-52},
	year = {2014},
	issn = {0167-9473},
	author = {A. Butler and R.D. Haynes and T.D. Humphries and P. Ranjan}
}

@article{toal2011,
	author = {D. J.J. Toal and N. W. Bressloff and A. J. Keane and C. M.E. Holden},
	title = {The development of a hybridized particle swarm for kriging hyperparameter tuning},
	journal = {Engineering Optimization},
	volume = {43},
	number = {6},
	pages = {675--699},
	year = {2011},
	publisher = {Taylor \& Francis}
}

@article{matthews2017,
	author = {Matthews, Alexander G. de G. and van der Wilk, Mark and Nickson, Tom
	and Fujii, Keisuke and Boukouvalas, Alexis and Le{\'o}n-Villagr{\'a}, Pablo
	and Ghahramani, Zoubin and Hensman, James},
	title   = "{GPflow: A Gaussian process library using TensorFlow}",
	journal = {Journal of Machine Learning Research},
	year    = {2017},
	volume  = {18},
	number  = {40},
	pages   = {1--6}
}

@article{binois2018,
	author = {Mickaël Binois and Robert B. Gramacy and Mike Ludkovski},
	title = "{Practical Heteroscedastic Gaussian Process Modeling for Large Simulation Experiments}",
	journal = {Journal of Computational and Graphical Statistics},
	volume = {27},
	number = {4},
	pages = {808--821},
	year = {2018},
	publisher = {Taylor \& Francis}
}

@article{gramacy2012,
	author = {Gramacy, Robert B. and Lee, Herbert K. H.},
	journal = {Statistics and Computing},
	number = {3},
	pages = {713--722},
	title = {Cases for the nugget in modeling computer experiments},
	volume = {22},
	year = {2012}
}
\bibliographystyle{plain}
\end{document}